\documentclass[conference]{IEEEtran}
\IEEEoverridecommandlockouts
\usepackage{cite}
\usepackage{amsmath,amssymb,amsfonts}
\usepackage{algorithmic}
\usepackage{graphicx}
\usepackage{textcomp}
\usepackage{xcolor}
\usepackage{url}

\def\BibTeX{{\rm B\kern-.05em{\sc i\kern-.025em b}\kern-.08em
    T\kern-.1667em\lower.7ex\hbox{E}\kern-.125emX}}
\begin{document}

\title{Low-Cost IoT-Enabled Tele-ECG Monitoring for Resource-Constrained Settings: System Design and Prototype \\
% {\footnotesize \textsuperscript{*}Note: Sub-titles are not captured in Xplore and
% should not be used}
\thanks{Special appreciation goes to Medi Clinic (Pvt. Ltd.)for their invaluable support in providing the ECG data essential for this research. The contributions of various researchers and authors in ECG monitoring, whose works provided a strong foundation for this study, are also acknowledged.}
}

\author{\IEEEauthorblockN{1\textsuperscript{st} Seemron Neupane }
\IEEEauthorblockA{\textit{Department of Electronics and Computer Engineering} \\
\textit{Pashchimanchal Campus, Tribhuvan University}\\
Pokhara, Nepal \\
seemronneupane@gmail.com}
\and
\IEEEauthorblockN{2\textsuperscript{nd} Aashish Ghimire}
\IEEEauthorblockA{\textit{Department of Computer Science} \\
\textit{University of South Dakota}\\
South Dakota, USA  \\
aashishg.ghimire@coyotes.usd.edu}
% \and
% \IEEEauthorblockN{3\textsuperscript{rd} Given Name Surname}
% \IEEEauthorblockA{\textit{dept. name of organization (of Aff.)} \\
% \textit{name of organization (of Aff.)}\\
% City, Country \\
% email address or ORCID}
% \and
% \IEEEauthorblockN{4\textsuperscript{th} Given Name Surname}
% \IEEEauthorblockA{\textit{dept. name of organization (of Aff.)} \\
% \textit{name of organization (of Aff.)}\\
% City, Country \\
% email address or ORCID}
% \and
% \IEEEauthorblockN{5\textsuperscript{th} Given Name Surname}
% \IEEEauthorblockA{\textit{dept. name of organization (of Aff.)} \\
% \textit{name of organization (of Aff.)}\\
% City, Country \\
% email address or ORCID}
% \and
% \IEEEauthorblockN{6\textsuperscript{th} Given Name Surname}
% \IEEEauthorblockA{\textit{dept. name of organization (of Aff.)} \\
% \textit{name of organization (of Aff.)}\\
% City, Country \\
% email address or ORCID}
}

\maketitle

\begin{abstract}
With the availability of automation machinery and its superiority, are being slothful and inviting many diseases to invade them. The world still has so many places where people lack basic health facilities. Due to early detection and intervention, CDV can be cured to an extreme extent. It heavily reduces travel and associated costs. A remote ECG monitoring system enables community health workers to support and empower patients through telemedicine. However, there remains some financial and logistical burden. Heart disease cannot be taken lightly. These patients require regular health check-ups and the attention of health personnel in a short period if their health deteriorates suddenly and rapidly. Chronic diseases are extremely variable in their symptoms and evolution of treatment. Some, if not treated early, will end the patient's life. The trend of the INTERNET OF THINGS, IoT, is spreading massively. This paper focuses on the three main: the operator, the doctor, and the server over which the data is being sent. 

\end{abstract}


\section{Introduction}
Heart disease, commonly referred to as cardiovascular disease (CVD), includes a group of conditions affecting the heart and blood vessels. Globally, CVD contributes to approximately 17.9 million deaths per year (about 32\% of total deaths) and remains a leading cause of mortality \cite{shrestha2021}. The burden varies across regions due to differences in lifestyle, climate, healthcare access, and socioeconomic conditions. For instance, the high-income Asia Pacific region reports comparatively lower CVD mortality, while Eastern Europe reports higher mortality \cite{shrestha2021}.

In Nepal, CVD is also a major cause of death, contributing to around 30\% of total deaths \cite{khanal2018}. Despite ongoing efforts, access to timely cardiac monitoring remains limited in many areas. A significant portion of the population continues to face barriers in receiving basic healthcare services, particularly in remote regions where patients may need to travel long distances for routine checkups and referrals. Existing reports highlight gaps in Nepal's health system capacity for managing chronic conditions such as heart disease and diabetes \cite{nmc2023}. In addition, the distribution of specialized clinicians is limited, and the country has a relatively small number of cardiologists compared to overall demand \cite{nmc2023}. The health sector allocation in fiscal year 23/24 (red book) indicates a constrained budget share for strengthening healthcare infrastructure and services, which reflects a common challenge across developing countries \cite{budget2024}.

These constraints motivate the development of low-cost and practical monitoring tools that can support remote follow-up and early attention when a patient's condition changes. With the rapid growth of the Internet of Things (IoT), interconnected devices equipped with sensors and network connectivity can transmit physiological data over distance with minimal human intervention. For cardiac assessment, electrocardiogram (ECG) monitoring measures the electrical activity generated during heart contraction. Electrodes placed on the body capture the signal, which can be visualized as a waveform for clinical interpretation. A remote ECG monitoring system can reduce travel burden, support community health workers, and enable clinicians to review both real-time and stored waveforms through telemedicine.

The system presented in this paper focuses on three operational entities: (i) the patient-side operator, (ii) the clinician/doctor, and (iii) a server that receives, stores, and serves ECG data for visualization. The design emphasizes affordability, ease of use, and an end-to-end pipeline from acquisition to mobile visualization. The \textbf{main contribution of the paper:}
\begin{itemize}
\item Development of a low-cost ECG acquisition module using AD8232 with ESP32-based wireless transmission.
\item Integration of a backend server (Django) to store ECG samples and support REST-based communication.
\item A Flutter-based mobile interface for continuous waveform visualization and remote access.
\item A modular architecture that can be extended for alerts and future analysis (e.g., anomaly detection) without redesigning the acquisition hardware.
\end{itemize}

\section{Literative review}
Concept reuse in monitoring systems is important for the development and improvement of remote ECG solutions. The iterative cycle begins with understanding functional requirements such as acquisition quality, real-time access, and usability for clinicians and patients. The book \textit{Internet of Things with ESP32} by Ruben Oliva Ramos and Agus Kurniawan provides practical guidance on building scalable low-power IoT devices and discusses protocol-level considerations that are relevant to healthcare monitoring \cite{ramos2018}. 

Adam Gacek and Witold Pedrycz bridge biomedical engineering and healthcare practice in their book \textit{ECG Signal Processing, Classification and Interpretation}, which discusses ECG signal characteristics and processing concepts that support hardware-based acquisition \cite{gacek2012}. 

S. Kumar and N. Sharma presented an IoT-based wearable ECG monitoring approach using AD8232 with ESP32 and cloud access, highlighting early detection potential and remote availability of ECG signals \cite{kumar2019}. Similarly, a wearable sensor-based approach for real-time cardiac monitoring with cloud computing and wireless communication is discussed in \textit{Remote Health Monitoring System for Detecting Cardiac Disorders}, which emphasizes reduced hospital visits for remote patients \cite{bansal2015}. 

Serhani et al. reviewed ECG monitoring architectures and key challenges, and highlighted the need for connected, efficient, and cost-effective systems using IoT and data-driven techniques \cite{hadeel2020}. Motivated by these studies, the proposed system in this paper focuses on a complete operational pipeline that connects sensing, transmission, storage, and mobile visualization under a lightweight implementation.

\section{Project Description}
The Remote ECG Monitoring System project aims to develop a practical end-to-end system that enables ECG acquisition, transmission, storage, and visualization. The objective is to support real-time observation of ECG waveforms and maintain stored records for follow-up and review. The system is designed to meet the typical needs of remote monitoring: portability at the patient side, reliable data delivery to a backend server, and accessible visualization at the clinician side.

The project has two sides of work:

i. Endside user (patient side)

ii. The doctor

\subsection{Hardware Components}

\begin{table}[h]
\centering
\caption{Major components and their roles in the proposed system.}
\renewcommand{\arraystretch}{1}
\begin{tabular}{|p{1.5cm}|p{1.5cm}|p{4cm}|}
\hline
\textbf{Module} & \textbf{Technology} & \textbf{Role in the system} \\
\hline
ECG front-end & AD8232 & Amplifies and conditions ECG biopotential signal \\
\hline
Controller \& radio & ESP32 & Samples analog ECG and transmits via Wi-Fi \\
\hline
Electrodes & RA/LA/RL leads & Captures ECG signal from skin surface \\
\hline
Backend server & Django & Stores samples and serves APIs for clients \\
\hline
Mobile app & Flutter & Real-time waveform visualization and patient view \\
\hline
\end{tabular}
\label{tab:components}
\end{table}

\subsubsection{ESP32}
ESP32 is a 32-bit microcontroller with integrated Wi-Fi and Bluetooth capability. Its wireless connectivity and low-power features support portable IoT designs and make it suitable for transmitting physiological signals such as ECG.

\subsubsection{AD8232 ECG SENSOR}
The AD8232 is an ECG sensor module used to acquire and condition cardiac electrical signals. It provides amplification and filtering of low-amplitude biopotentials and produces an analog output waveform. Since ECG signals are commonly affected by noise and motion artifacts, the AD8232 front-end helps obtain a cleaner waveform for reliable visualization and transmission.

\subsubsection{AD8232 PIN CONFIGURATION}
There are nine connecting pins and wires on the AD8232 operational amplifier. The monitoring module includes pins such as SDN, LO+ pin, LO- pin, OUTPUT, 3.3V, and GND. The module also supports 3-lead electrodes: RA (Right Arm), LA (Left Arm), and RL (Right Leg). The electrodes are placed at appropriate spots on the body to obtain the ECG signal, and the AD8232 sensor is connected to the ESP32 development kit.

\subsubsection{ELECTRODES}
Disposable electrodes and lead connectors are used to establish reliable contact with the skin. Proper electrical and mechanical coupling improves signal stability and reduces artifacts caused by loose connections. The connector design helps ease attachment and removal while maintaining adequate contact for ECG acquisition.

\begin{figure}[h]
    \centering
    \includegraphics[width=\columnwidth]{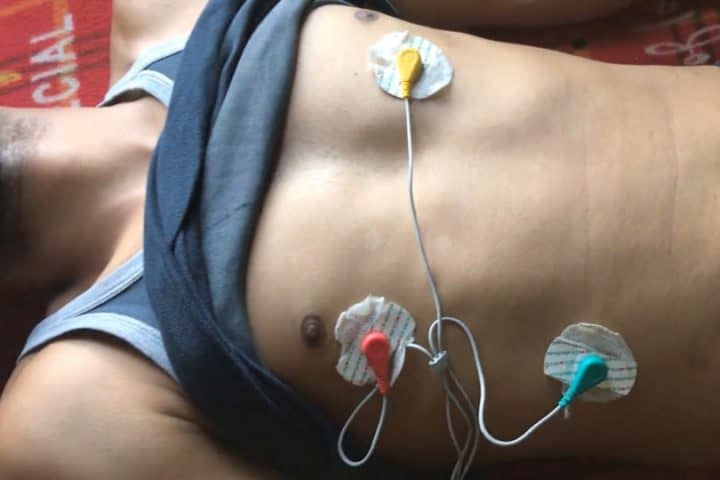} % Change the filename and format accordingly
    \caption{Electrode placement}
    \label{fig:1} % Label for referencing in the text
\end{figure}

\section{Methodology}
In underdeveloped and developing contexts, it may not be convenient for a patient with cardiovascular issues to monitor heart activity regularly through hospital visits. Traditional ECG machines are often bulky, relatively expensive, and not intended for continuous home-based monitoring. The proposed approach uses disposable electrodes and a compact acquisition module to capture ECG signals and transmit them wirelessly for remote observation. For improved signal quality, the skin is kept dry and electrode placement is performed carefully as shown in Fig. \ref{fig:1} to reduce noise and obtain a stable waveform.

The heart rhythm is captured via electrodes attached to the AD8232 module. The conditioned analog signal is sampled by ESP32, and data is transmitted through Wi-Fi to a server for visualization and storage. ESP32 is used for acquiring and transmitting the ECG signal, and the power supply can be provided through battery or USB. Power optimization can be further supported using deep sleep modes and low-power management strategies. For the software part, Arduino IDE is used for ESP32 firmware development and HTTP transmission to a cloud-based system. Flutter is used for mobile application development, and the application receives ECG data through the backend. The backend server is implemented in Django, which stores patient data and supports real-time waveform visualization through API communication.

\begin{table}[h]
\centering
\caption{Design requirements and how the proposed system meets them.}
\renewcommand{\arraystretch}{1}
\begin{tabular}{|p{1.8cm}|p{1.5cm}|p{4.0cm}|}
\hline
\textbf{Design requirement} & \textbf{Target} & \textbf{How it is addressed} \\
\hline
Portability & Compact setup & AD8232 + ESP32 with disposable electrodes \\
\hline
Low cost & Affordable BOM & Uses widely available modules and components \\
\hline
Remote accessibility & Networked monitoring & Wi-Fi transmission to server and mobile app \\
\hline
Ease of use & Simple operation & Minimal steps: place leads, power device, view app \\
\hline
Extensibility & Future features & Backend supports adding alerts/analysis modules \\
\hline
\end{tabular}
\label{tab:requirements}
\end{table}

\section{Interfacing and Working}

\begin{figure}[h]
    \centering
    \includegraphics[width=0.5\textwidth]{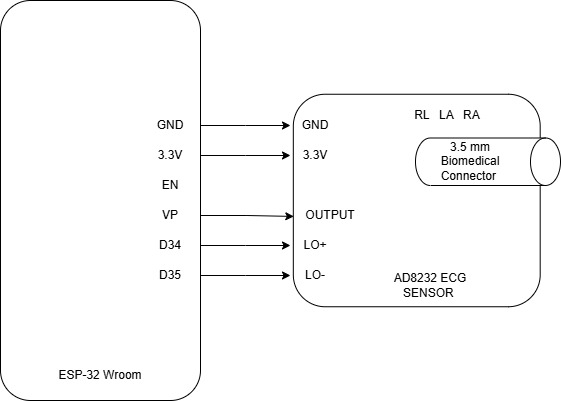} 
    \caption{System block diagram}
    \label{fig:2} % Label for referencing in the text
\end{figure}

\begin{table}[h]
\centering
\caption{End-to-end data flow in the remote ECG monitoring pipeline.}
\renewcommand{\arraystretch}{1}
\begin{tabular}{|p{1.3cm}|p{2.2cm}|p{4cm}|}
\hline
\textbf{Stage} & \textbf{Input/Output} & \textbf{Operation} \\
\hline
Acquisition & Electrodes $\rightarrow$ AD8232 & Captures ECG and performs analog conditioning \\
\hline
Sampling & AD8232 OUTPUT $\rightarrow$ ESP32 VP & Converts analog waveform to digital samples \\
\hline
Transmission & ESP32 $\rightarrow$ Server & Sends samples using HTTP over Wi-Fi \\
\hline
Storage & Server $\rightarrow$ Database & Stores ECG samples for retrieval and review \\
\hline
Visualization & Database/API $\rightarrow$ Flutter app & Displays real-time waveform and history \\
\hline
\end{tabular}
\label{tab:pipeline}
\end{table}

At every beat, the heart is depolarized to trigger contraction, and the electrical activity propagates through the body and can be detected at the skin surface. This is the principle behind ECG measurement. An ECG machine records this activity via electrodes on the skin and displays it graphically. In the proposed system, Fig. \ref{fig:2} illustrates the overall workflow where AD8232 performs signal conditioning, ESP32 handles sampling and Wi-Fi transmission, and the server provides storage and visualization through the mobile application.

\begin{table}[h]
    \centering
        \caption{Pin Connection}
    \renewcommand{\arraystretch}{1} % Adjust row spacing
    \begin{tabular}{|p{2cm}|p{2cm}|p{3.5cm}|}
        \hline
        \textbf{AD8232 Pin} & \textbf{Connected to ESP32} & \textbf{Description} \\
        \hline
        OUTPUT & VP & Outputs the ECG analog signal \\
        \hline
        3.3V & 3.3V & Powers the AD8232 sensor \\
        \hline
        GND & GND & Ground connection \\
        \hline
        L0+ & D34 & Detects if the positive electrode is detached \\
        \hline
        LO- & D35 & Detects if the negative electrode is detached \\
        \hline
        SDN & Not Connected & Used for shutdown mode (optional) \\
        \hline
    \end{tabular}
    \label{tab:sample}
\end{table}

The ESP32 controller operates at 3.3V and supplies power to the AD8232 module. Electrodes placed on the patient's body capture the ECG waveform as an analog biopotential signal. AD8232 amplifies and filters this signal and provides an analog output to the ESP32 VP pin for sampling. Lead-off pins LO+ and LO- provide electrode detachment information, which can be used to improve reliability and reduce false readings due to poor contact.

After acquisition, the sampled ECG data is transmitted to the backend server. Django stores the received samples in a database so that ECG data can be accessed for later review. The mobile application fetches waveform samples through the backend and displays the live ECG signal. This end-to-end integration enables real-time monitoring and also supports storing historical ECG records. The backend design is kept modular so that additional features such as automated alerts and anomaly detection can be integrated in future versions.

\section{Result}

\begin{figure}[h]
    \centering
    \includegraphics[width=0.4\textwidth]{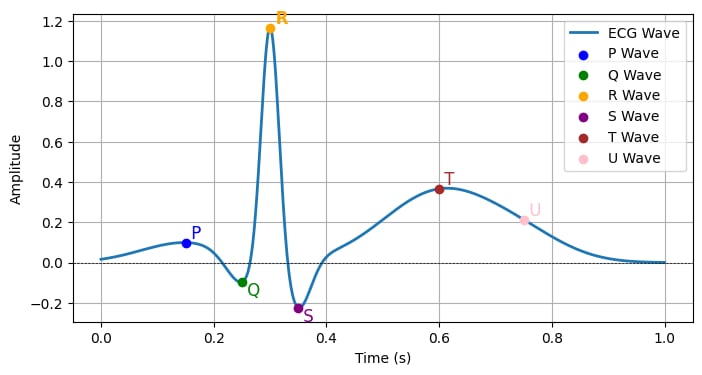}  % Replace with your image file name
    \caption{IoT extracted PQRST cycle}
    \label{fig 3:example}
\end{figure}

The PQRST waveform obtained from the proposed system is shown in Fig. \ref{fig 3:example}. The system successfully captured ECG signals from the AD8232 sensor and transmitted the waveform to the Django backend over Wi-Fi. The received data was stored in the database, allowing both real-time visualization and future access. The mobile application provided continuous waveform display and served as the interface between the patient-side device and medical personnel. The observed waveform morphology demonstrates that the acquisition and transmission pipeline can deliver an interpretable ECG trace suitable for remote monitoring. The current backend can also be extended to include additional functions such as signal quality assessment, alert generation, and machine learning based anomaly detection.

\section{Conclusion}
The device designed and implemented in this work demonstrates a low-cost IoT-based remote ECG monitoring system suitable for practical deployment. By integrating AD8232-based ECG acquisition, ESP32 wireless transmission, and a Django--Flutter software stack, the system supports real-time waveform visualization and database storage for follow-up. The architecture reduces dependence on frequent hospital visits and provides a foundation for telemedicine-based monitoring in resource-constrained settings. The modular server design also makes it feasible to incorporate advanced analysis and alert mechanisms in later stages.

\section{Future Enhancement}
Future enhancements will focus on improving system reliability, security, and clinical utility. This includes implementing basic signal quality checks, lead-off based alerts, and secure communication mechanisms for protecting patient data. In addition, automated classification of abnormal ECG patterns and anomaly detection modules can be integrated at the backend to support early screening. Long-term extensions may include clinician dashboards for trend analysis and improved workflow integration for remote follow-up.

\bibliographystyle{unsrt}
\bibliography{refs}

@article{shrestha2021,
  author  = {Shrestha, A. and Maharjan, R. and Karmacharya, B. M.},
  title   = {Health System Gaps in Cardiovascular Disease Prevention and Management in Nepal},
  journal = {BMC Health Services Research},
  volume  = {21},
  number  = {1},
  pages   = {2--13},
  year    = {2021}
}

@article{khanal2018,
  author  = {Khanal, M. K. and Ahmed, M. S. A. M. and others},
  title   = {Prevalence and Clustering of Cardiovascular Disease Risk Factors in Rural Nepalese Population Aged 40--80 Years},
  journal = {BMC Public Health},
  volume  = {18},
  number  = {1},
  year    = {2018}
}

@misc{nmc2023,
  title        = {Specialist Doctors in Nepal Crosses 10 Thousand: Nepal Medical Council (NMC)},
  howpublished = {[Online]. Available: \url{https://www.collegenp.com/news/specialist-doctors-in-nepal-crosses-10-thousand-nmc}},
  year         = {2023},
  month        = feb
}

@misc{budget2024,
  title        = {Government Hikes Health Sector Budget by Rs 14 Billion in FY 2023/24},
  howpublished = {[Online]. Available: \url{https://myrepublica.nagariknetwork.com/news/govt-hikes-health-sector-budget-by-rs-14-billion-in-fy-2023-24/}},
  year         = {2024},
  month        = may
}

@book{ramos2018,
  author    = {Ramos, R. O. and Kurniawan, A.},
  title     = {Internet of Things with ESP32},
  publisher = {Packt Publishing},
  address   = {Birmingham, U.K.},
  year      = {2018}
}

@book{gacek2012,
  editor    = {Gacek, A. and Pedrycz, W.},
  title     = {ECG Signal Processing, Classification and Interpretation},
  publisher = {Springer},
  address   = {London, U.K.},
  year      = {2012}
}

@inproceedings{kumar2019,
  author    = {Kumar, S. and Singh, S. P. and Sharma, N.},
  title     = {IoT Based Portable ECG Monitoring Device for Smart Healthcare},
  booktitle = {Proc. 6th Int. Conf. Signal Processing and Integrated Networks (SPIN)},
  pages     = {327--331},
  year      = {2019}
}

@article{bansal2015,
  author  = {Bansal, A. and Kumar, S. and Baipai, A.},
  title   = {Remote Health Monitoring System for Detecting Cardiac Disorders},
  journal = {IET Systems Biology},
  volume  = {9},
  number  = {6},
  pages   = {309--314},
  year    = {2015}
}

@article{hadeel2020,
  author  = {Hadeel, M. A. S. and Ismail, T. E. K. H. and Navaz, A. N.},
  title   = {ECG Monitoring Systems: Review, Architecture, Processes and Key Challenges},
  journal = {Sensors},
  volume  = {20},
  number  = {6},
  year    = {2020}
}

\vspace{12pt}

\end{document}